\documentclass[twocolumn,showpacs,letterpaper,showpacs,superscriptaddress]{revtex4}
\usepackage{graphicx,amsmath,amssymb,amsfonts,latexsym,color,dcolumn,bm,subfigure}

\renewcommand{\figurename}{{\bf Figure}}

\begin{document}

\title{Strong Casimir force reduction through metallic surface nanostructuring}

\author{Francesco Intravaia}
\affiliation{Theoretical Division, MS B213, Los Alamos National Laboratory, Los Alamos, New Mexico 87545, USA}
\author{Stephan Koev}
\affiliation{Center for Nanoscale Science and Technology, National Institute of Standards and Technology, Gaithersburg, Maryland 20899, USA}
\affiliation{Maryland Nanocenter, University of Maryland, College Park, MD 20742, USA}
\author{Il Woong Jung}
\affiliation{Center for Nanoscale Materials, Argonne National Laboratory, Argonne, Illinois 60439, USA }
\author{A. Alec Talin}
\affiliation{Center for Nanoscale Science and Technology, National Institute of Standards and Technology, Gaithersburg, Maryland 20899, USA}
\author{Paul S. Davids}
\affiliation{Applied Photonics and Microsystems, Sandia National Laboratories, Albuquerque, New Mexico 87185, USA}
\author{Ricardo S. Decca}
\affiliation{Department of Physics, Indiana University-Purdue University Indianapolis, Indianapolis, Indiana 46202, USA}
\author{Vladimir A. Aksyuk}
\affiliation{Center for Nanoscale Science and Technology, National Institute of Standards and Technology, Gaithersburg, Maryland 20899, USA}
\author{Diego A. R. Dalvit}
\affiliation{Theoretical Division, MS B213, Los Alamos National Laboratory, Los Alamos, New Mexico 87545, USA}
\author{Daniel L\'{o}pez}
\affiliation{Center for Nanoscale Materials, Argonne National Laboratory, Argonne, Illinois 60439, USA }

\date{ \today}

\begin{abstract}
The Casimir force between bodies in vacuum can be understood as arising from their interaction with an infinite number of fluctuating electromagnetic quantum vacuum modes, resulting in a complex dependence on the shape and material of the interacting objects. Becoming dominant at small separations, the force plays a significant role in nanomechanics and object manipulation at the nanoscale, leading to a considerable interest in identifying structures where the Casimir interaction behaves significantly different from the well-known attractive force between parallel plates. Here we experimentally demonstrate that by nanostructuring one of the interacting metal surfaces at scales below the plasma wavelength, an unexpected regime in the Casimir force can be observed. Replacing a flat surface with a deep metallic lamellar grating with sub-100 nm features strongly suppresses the Casimir force and for large inter-surfaces separations reduces it beyond what would be expected by any existing theoretical prediction.

\end{abstract}

\maketitle

The Casimir effect, in its most basic form, can be understood as a direct macroscopic manifestation of quantum electrodynamics, whereby changing the relative position of metallic or dielectric bodies modifies the zero point energy of the surrounding electromagnetic vacuum, resulting in a measurable interaction force between them \cite{1}. This direct connection to fundamental concepts in quantum mechanics has made this effect the object of continuous theoretical and experimental attention for over 60 years since it was first brought to light by H. Casimir. More broadly, it is also a particular case of fluctuation-induced interaction phenomena encountered in a wide variety of physical systems, such as binary liquid mixtures \cite{2}, cell membranes and proteins \cite{3}, and even in cosmology \cite{4}. 

The Casimir force has important technological consequences and untapped application potential in the field of micro- and nano-electromechanical systems - engineered devices with moveable parts ranging from 500 $\mu$m down to 10 nm in size. 
For example, in nanoelectromechanical contact switches \cite{4bis} currently being developed as complements or even potential successors of conventional CMOS, movable parts are separated by much less than 1 $\mu$m, and accounting for the Casimir force is essential for their correct design and functioning. It has been shown that this force significantly modifies both the static and the dynamic MEMS performance, leads to unwanted stiction, and is an important source of nonlinear behavior \cite{5}. On the other hand, it has potential uses for non-contact low-dissipation actuation and tuneability of such nanomachines \cite{6}. Beyond nanomechanics, controlling this force is important for a diversity of fields, ranging from quantum computing with atom chips \cite{7} to searches for non-Newtonian gravity at sub-micron scales \cite{8,8'}.

The seminal theoretical work of Lifshitz \cite{9} on the Casimir force between planar closely spaced dielectric surfaces led to a complete framework for computing forces arising from fluctuating electromagnetic fields. Well-established approximations, such as the proximity force approximation \cite{10} (PFA), have been widely utilized to extend the theory to non-planar complex geometries. The PFA assumes that the force between non-planar objects is the sum of the forces between infinitesimal planar sections computed with Lifshitz's approach. The theory has been experimentally verified under a broad range of conditions, e.g., at different length scales, where either quantum or thermal fluctuations dominate the interaction, with different materials and even with fluids between the surfaces \cite{11,12,13,14,15,16,17,18,19}. However, with few exceptions \cite{20,21}, these precision measurements so far have been limited to planar or near-planar surfaces. The Casimir effect with complex, non-planar geometries, where simple approximations are not applicable, continues to present theoretical and experimental challenges. While the underlying theoretical principles, approaches, and approximations describing the interaction of electromagnetic waves with metallic and dielectric structures of complex shapes are well established in classical photonics, the extra challenge stems from the inherently broadband nature of the Casimir effect, where fluctuations at all frequencies and wave-vectors have to be taken into account simultaneously. This not only makes the problem more complex and less amenable to an analytical solution, but also many of the abstractions based on narrow-band intuition become less applicable.

In the last few years, advances in numerical techniques give us the tools to compute the Casimir force between complex structures made of real materials \cite{22,23,24}. However, on the experimental side, due to the difficulties associated with the reliable fabrication of nanostructured samples and the measurement of the force, there have been very few measurements involving nanostructured surfaces. Only recently the Casimir interaction between nanostructured silicon gratings and a gold-coated sphere has been measured, with conclusive evidence of the strong geometry dependence and non-additivity of the Casimir force \cite{20}. Specifically, it was observed that the patterning of periodic nanoscale trenches into a silicon substrate makes the Casimir force per unit area more attractive than the corresponding PFA prediction. 

Metallic nanostructures have the potential to unveil a new realm for Casimir force manipulation. Indeed, they support collective surface EM modes called surface plasmons, which can propagate along the surface, decay exponentially away from it, and have a characteristic frequency of the order of the plasma frequency. In the simple plane-plane configuration, it is known that surface plasmons affect the Casimir force in a non-trivial manner, featuring an attractive (repulsive) contribution to the force for distances shorter (larger) than the plasma wavelength \cite{25}. Nanostructured surfaces with tailored plasmonic dispersion relations have already impacted classical nano-photonics, with applications ranging from extraordinary light transmission to surface-enhanced Raman scattering \cite{26,27}. Likewise, metallic structures, with strong deviations from the planar geometry and possessing geometrical features on very small scales, are likely to give significant new insights into potential Casimir devices. Here we investigate the impact of nanostructuring one of the interacting metal surfaces at scales below the plasma wavelenght. We have designed and fabricated high aspect ratio nanostructured gold gratings with critical dimensions ranging from 90 nm to 200 nm, and we have then performed high precision measurements of the Casimir force in vacuum between a gold-coated sphere and the nanostructured gold gratings. A new regime in the Casimir interaction has been observed, significantly different from the well-known attraction between parallel plates.

\section*{Results}

\begin{figure}[t]
\includegraphics[width=\columnwidth]{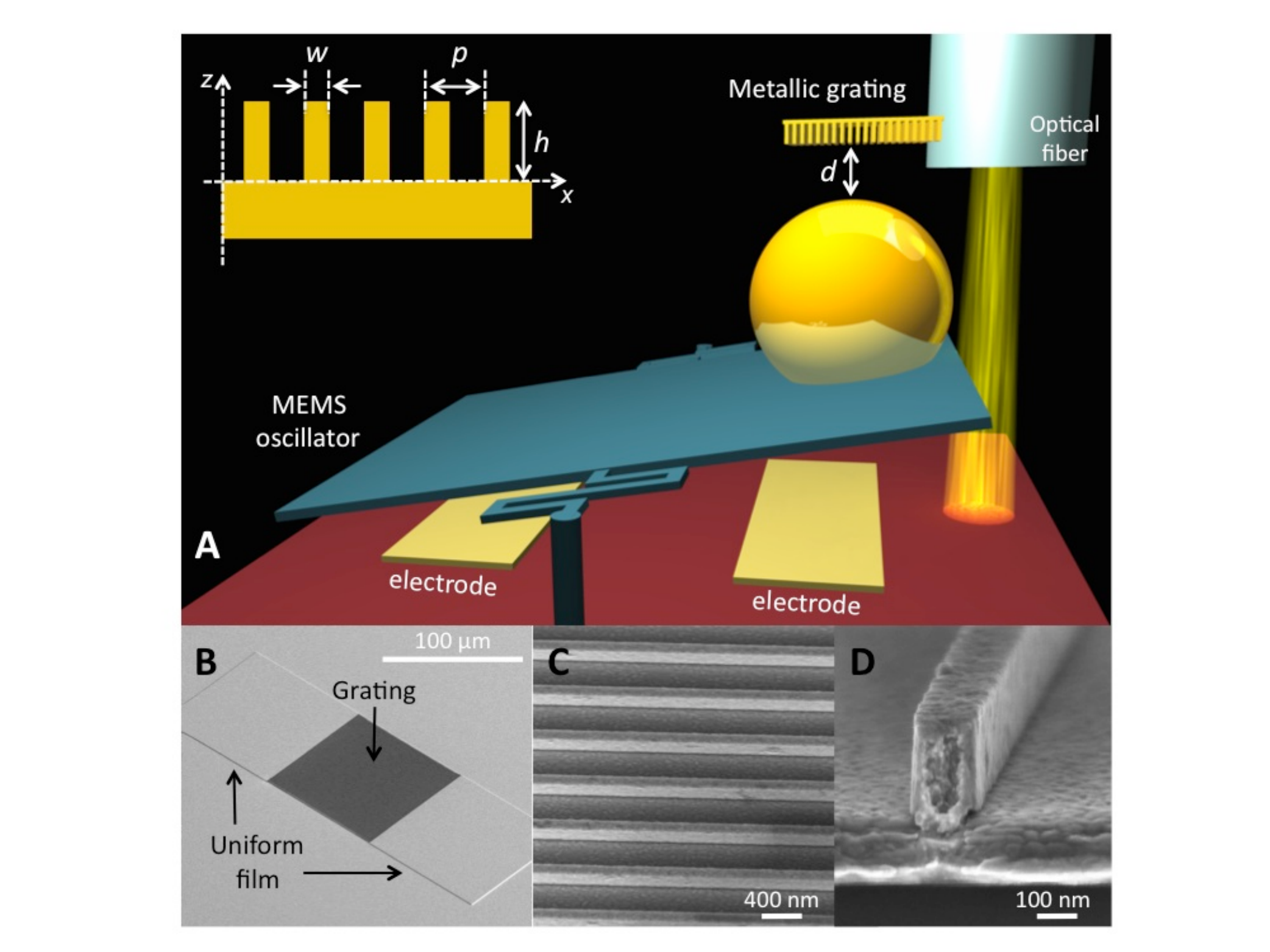}
\caption{{\bf Experimental configuration and sample details.}
{\bf (A)} Schematic drawing of the experimental configuration used to measure the Casimir force between a gold-coated sphere and a nanostructured grating. The sphere is attached to the torsional plate of a micromechanical oscillator and the nanostructured grating is fixed to a single mode optical fiber. The optical fiber is used to monitor the distance between the bottom of the fiber and the supporting substrate while the micromechanical oscillator provided a capacitive measurement of the Casimir interaction. (Inset) Definitions of the geometrical parameters of the metallic nanostructures. {\bf (B-D)} SEM images of typical samples used in the reported experiments.  {\bf (B)} The nanostructured gratings are limited by two uniform films used for calibration and reference(scale bar is 100 $\mu$m). {\bf (C)} Magnified detail of the grating area showing the high spatial uniformity achieved in these samples (scale bar, 400 nm). {\bf (D)} SEM cross-sectional photograph of a single grating element (scale bar, 100 nm).}
\label{Figure1}
\end{figure}

\noindent{\bf Sample Fabrication.} Fabrication of large area Au nanostructures with uniform sub-100 nm in-plane dimensions and vertical sidewalls with depth $\geq$ 200 nm (aspect ratio larger than one) remains challenging. The standard dry etching techniques, e.g. reactive ion etching, do not generally work well for noble metals because they generate large amount of sputtered material that re-deposit onto the structures being fabricated. Focused ion beam or ion milling may provide an alternative, but even in this case avoiding re-sputtering, creating deep vertical sidewalls and uniform depth, and ensuring the metal on the surface is pristine is close to impossible. Techniques based on gold deposition, e.g. lift-off or sputtering, are very popular but their applicability is limited to nanostructures with small aspect ratio ($<1$), while achieving narrow structures with tall vertical sidewalls and uniform height is limited by the inability to uniformly fill deep trenches during deposition. We were able to achieve this by first using state-of-the-art high-voltage 100 keV electron beam lithography to generate structures of the needed size and geometry in e-beam resist materials, and then using them as templates for pure gold deposition by either electroplating or sputtering. The e-beam parameters were carefully optimized to achieve the needed aspect ratios, separately for trenches in a positive tone resist for electroplating, and for ridges in a negative tone resist for sputtering. The gold deposition processes were also highly optimized to achieve uniform plating thickness and highly conformal sputtering coverage, both with low surface roughness. Figs. 1B to 1D show scanning electron microscopy (SEM) images with details of the gratings used.  The gratings dimensions are: width $w$ from 90 nm to 200 nm, period $p$ from 250 nm to 800 nm, and height $h$ from 200 nm to 500 nm.  A typical sample layout is shown in Fig. 1B. The grating area has dimensions of $50\times 50\; \mu{\rm m}^2$ and is surrounded by flat uniform gold films used for reference and calibration (see Methods).  Fig. 1C shows the uniformity of the nanofabricated surfaces and a cross-section of a single grating element is shown in Fig. 1D.  (See Methods for details on surface roughness and uniformity.) \\

\noindent{\bf Force measurement.} The experimental setup for measuring the Casimir effect is similar to the one we have used in previous work \cite{28}, which allowed us to perform the most precise measurement to date of the force between metallic surfaces. Fig. 1A shows a schematic of the experimental system used (see details in Methods). It consists of a metal-coated sphere of radius $R = (151.7 \pm 0.2) \; \mu$m, attached to a micro-mechanical torsional oscillator. The metallic grating is attached to an optical fiber that, at each distance $d$, is used to keep the sphere-grating separation stable within half a nanometer. As the grating is brought into close proximity of the sphere, the interaction between the two surfaces produces a shift in the oscillator resonance frequency, which is used to extract the gradient of the Casimir force,  $\partial_d F_\mathrm{{sg}}$. 

The measured force gradient shows finite size effects when the sphere is near the edge of the gratings. In order to avoid these effects, the sphere was positioned close to the center of the samples under investigation. The measurement results become experimentally indistinguishable from each other once the center of the sphere is at a distance of the order of or larger than 10 $\mu$m from the edge of the sample (See Methods).

The apparatus was calibrated using a known, calculable interaction - the electrostatic one. Two different calibration techniques were employed: the first one used the flat continuous film in the immediate vicinity of the grating for calibration just before performing any measurement on top of the grating area; the second one performed the whole electrostatic calibration process over the grating itself. The latter technique required calculating the system capacitance, which was done by solving the electrostatic problem using a finite elements analysis. Both procedures produced experimentally indistinguishable results (see Methods and Supplementary Figure S2). The use of a sphere instead of another planar surface avoids the problem of keeping the two objects parallel but complicates the exact theoretical description. A common approach to bypass this difficulty relies on the PFA: this approximation assumes that when the sphere's radius is much larger than the sphere-grating distance $d$ ($d/R\ll1$), the relevant EM field modes see the sphere as effectively planar and one can then approximate the sphere's surface as a collection of planar elements. Within this procedure the force gradient can be calculated as the sum of several local parallel plane interactions, which can be evaluated using the Lifshitz formula. Each local pressure depends on the distance between the sphere and the grating surface at that location, giving as a result 
\begin{eqnarray}
\partial_d F_\mathrm{{sg}}^{\rm PFA}(d) &=& 2 \pi R [ f P_\mathrm{{pp}}(d)+(1-f) P_\mathrm{{pp}}(d+h)] \nonumber \\
& \equiv & 2 \pi R P_\mathrm{{pg}}^{\rm PFA}(d) ,
\label{equation1}
\end{eqnarray}
where  $P_{\mathrm{pp}}(d)$ is the Lifshitz formula for the Casimir pressure between two parallel planes \cite{9}. In the previous expression the impact of nanostructuring is captured in the grating's filling factor $f=w/p$, and in the limit of $f \rightarrow 1$  we recover the usual result for the sphere-plane configuration.

\begin{figure}[t]
\includegraphics[width=1\columnwidth]{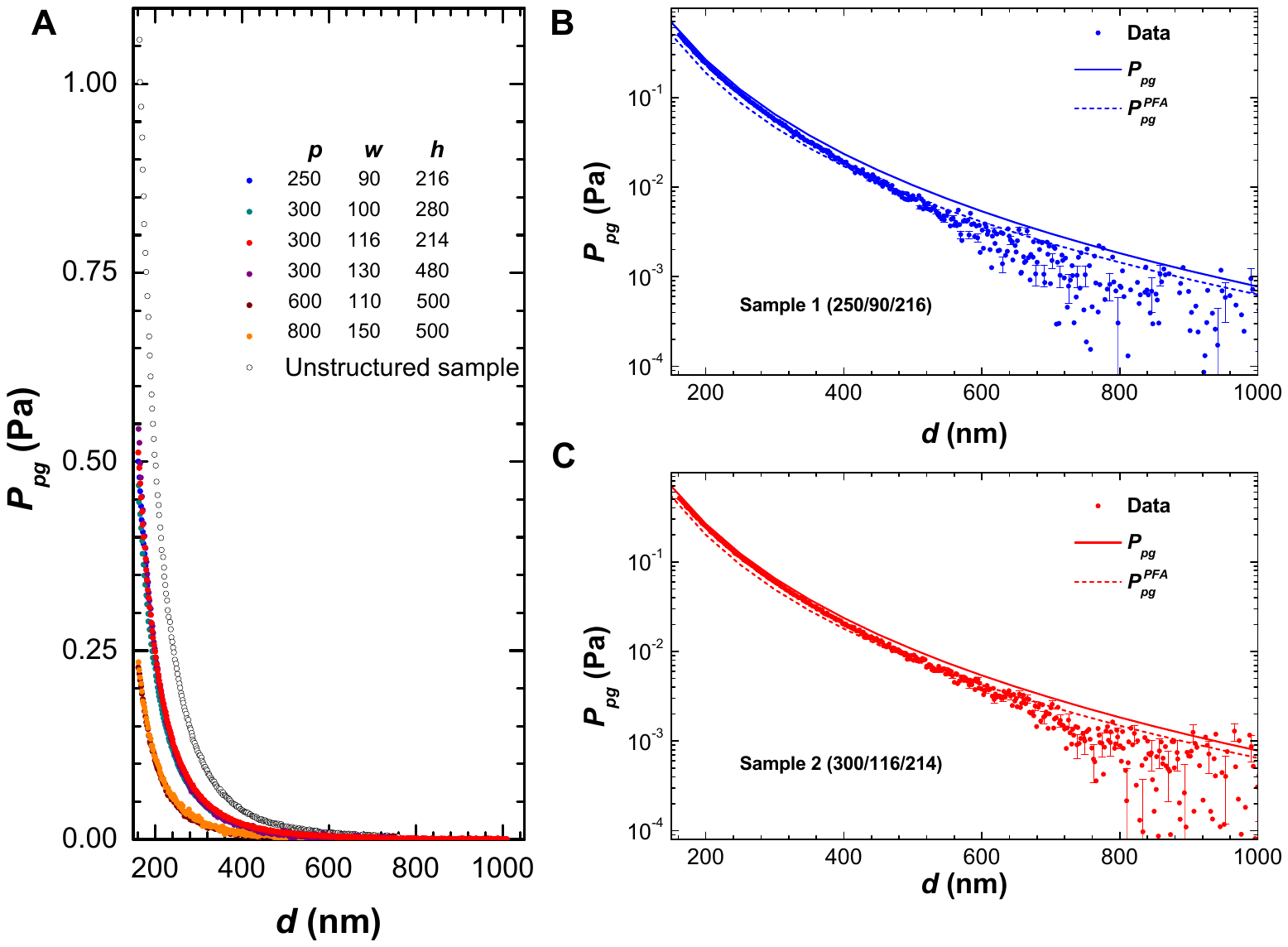}
\caption{{\bf Casimir pressure between the sphere and the gratings as a function of their separation.} {\bf (A)} Measurements done for metallic nanostructured samples with different parameters. The data show that the main effect of the nanostructure is to reduce the values of the pressure according to the samples� filling factors. {\bf (B,C)} Results for two samples of similar filling factors ($f_1=0.360$  and $f_2=0.387$, respectively). Experimental measurements (dots with error bars), proximity force approximation as in Eq. (\ref{equation1}) (dashed lines), and modal approach calculation as in Eq. (\ref{equation2}) (solid lines). Error bars are the variance of the mean measured pressure over the 45 repetitions of the experiment for each sample. They are plotted every fifth data point to increase the clarity of the figure (see Methods for more details). Geometrical parameters of the gratings are indicated as $p$ (period)/$w$ (width)/$h$ (height), all in nanometers.}
\label{Figure2}
\end{figure}

Fig. 2A shows the Casimir pressure for several gold nanostructured gratings with different dimensions. To simplify the data analysis, it is convenient to normalize the experimentally measured sphere-grating Casimir force gradient, $\partial_d F_\mathrm{{sg}}$, dividing it by the factor $2 \pi R$. Within PFA this ratio represents the data in terms of the equivalent plane-grating pressure $P_\mathrm{{pg}}$. 
As expected, as the filling factor is reduced the Casimir pressure is also reduced. This is a simple effect showing the dependence of the Casimir effect on the optical density of the involved bodies: nanostructuring leads indeed to a more diluted optical permittivity, which implies less force. In addition, the Casimir interaction for samples with similar filling factors appears to be independent of the height of the grating. For samples of similar geometry the results are substantially the same and do not depend on the sample preparation methodology, i.e. sputtering or electroplating. In order to clearly identify the influence of the grating geometry on the Casimir interaction, it is helpful to compare data obtained in nanostructured gratings having similar filling factors for which PFA predicts about the same result. Figures 2B and 2C show the data for two specific electroplated samples we will focus on in the remainder of the paper. Their filling factors are  $f_1=0.360$ (Fig. 2B) and $f_2=0.387$  (Fig. 2C).  The dashed lines show the behavior of the plane-grating pressure as calculated within a PFA treatment. It is clear from Figs. 2B and 2C that, in our case, the PFA gives a poor description of the equivalent plane-grating Casimir pressure. 

This disagreement is even more evident in Fig. 3 where the experimental data are normalized by the corresponding PFA expressions. By performing this normalization one suppresses geometrical effects associated with the filling factor and with the redefinition of the distance due to the height of the grating. As is clear in Fig. 3, this normalization gives very different results even for samples with similar filling factors. At short separations the experimental data show an equivalent pressure larger than the one predicted by PFA in Eq.(\ref{equation1}), i.e., the force per unit area becomes more attractive, similarly to what has been observed in silicon gratings \cite{20}. However, at large separations, the equivalent Casimir pressure is reduced with respect to the PFA prediction, i.e., the force per unit area becomes less attractive, reaching values more than 2 times smaller than the predicted by PFA. The separation at which the crossover between these two regimes occurs is roughly proportional to the period of the grating. This is the first experimental report of such significant pressure reduction with respect to the PFA prediction and of a crossover from enhancement to reduction of the Casimir force per unit area. The same behavior has been observed in each one of the 17 metallic gratings measured during this experiment, independently of the method used to fabricate the gratings. The inset in Fig. 3 shows the ratio $P_\mathrm{{pg}}/P_\mathrm{{pg}}^{\rm PFA}$ for a grating fabricated using sputtering techniques.

\begin{figure}[t]
\includegraphics[width=\columnwidth]{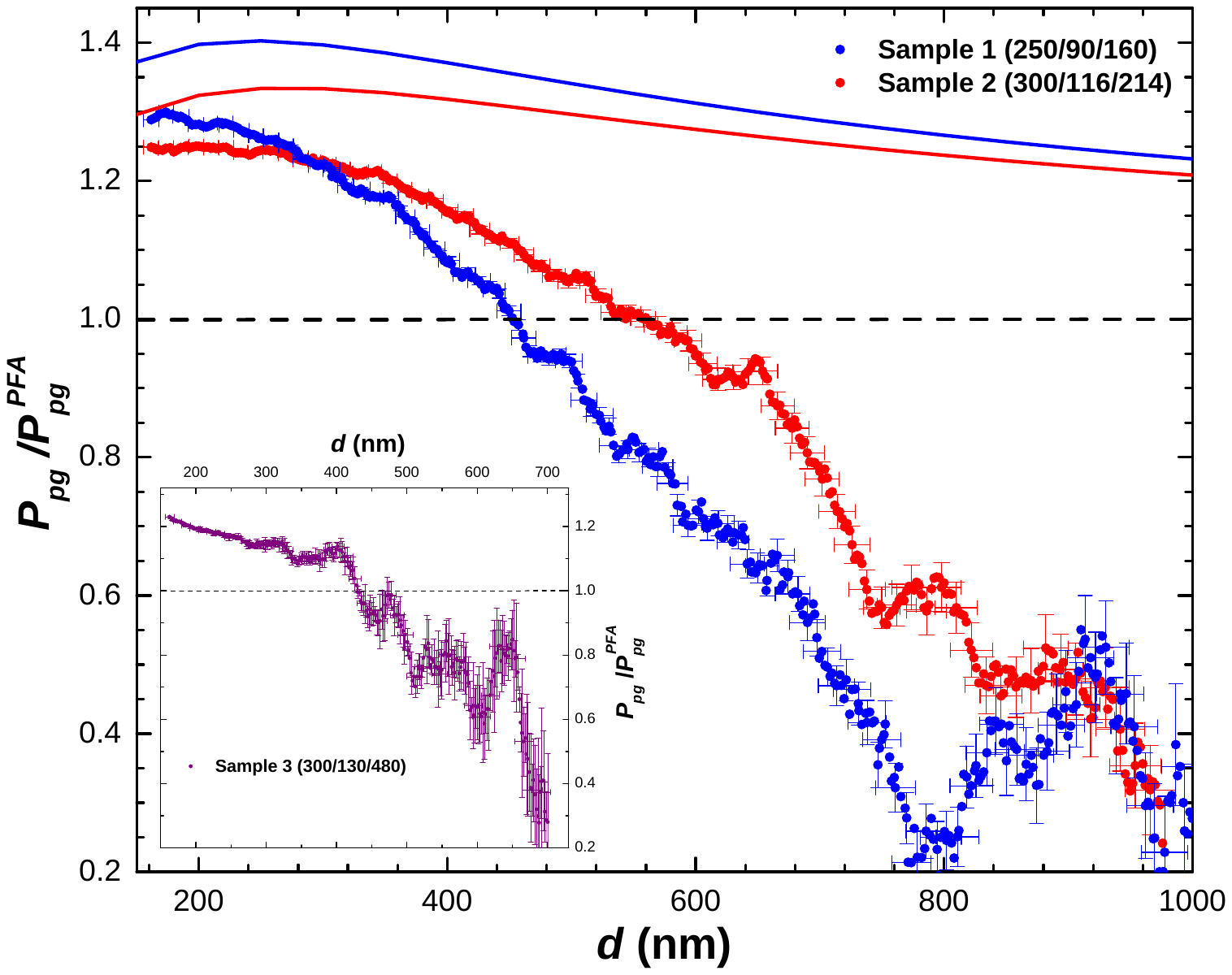}
\caption{{\bf Normalized Casimir pressure}. Main figure: Equivalent plane-grating Casimir pressure normalized by the PFA expression in Eq. (1), as a function of separation between the sphere and the grating for samples made with electroplating of high aspect ratio molds. Data are presented with dots and error bars for sample 1 (blue) and sample 2 (red). A weighted rolling average over a variable bin width is performed for the pressure, as described in the Methods section, which more details on error bars. The full lines correspond to the modal approach numerical prediction given in Eq. (2). Parameters are the same of Figure 2. 
The observed fluctuations on the experimental data originate from the rolling average. The effect of points far from the mean is extended over the bin where the average is performed.
Inset: Characteristic results for samples made by sputtering of Au onto hydrogen silesquioxane structures  (see Methods).}
\label{Figure3}
\end{figure}

\section*{Discussion}
An apparent trend is also clear when comparing data of nanostructures with similar filling factors but different periods: at short distances, the shorter the grating period the larger the enhancement of the Casimir pressure with respect to the PFA, while at large distances, the opposite happens - shorter period leads to a stronger reduction of the Casimir force. Since at large distances $\partial_d F_\mathrm{{sg}}^{\rm PFA}(d) \propto d^{-4}$, the data shows that the force gradient or the equivalent Casimir pressure decreases faster than the inverse fourth-power distance. In addition, for fixed filling factor the rate of decrease is larger for smaller period gratings. This latter observation can be understood from a scaling argument (see Supplementary Information). At these large distances, this behavior is in contradiction with what is intuitively expected: in the limit of small period one should progressively recover the plane-plane result. 

Deviations from PFA are not surprising and are known, at least theoretically, for some simple geometries. For example, in the sphere-plane configuration PFA overestimates  the exact result, the leading correction to PFA being of the order $d/R$ \cite{29,29'}. In the case of a doped silicon grating with micrometer features interacting with a Au sphere it was shown that PFA instead underestimates the correct result for plane-grating distances below 500 nm \cite{20}. To improve on our PFA expression (1) we numerically computed the equivalent plane-grating pressure $P_\mathrm{{pg}}$ for our configuration taking into account the small size and the high conductivity of our nanostructures, and approximated the sphere-grating force gradient in this modified PFA as $2 \pi R P_\mathrm{{pg}}(d)$. Within the scattering formalism \cite{23,24} the pressure is expressed as a series over Matsubara frequencies  $\xi_l=2 \pi l k_B T/\hbar$
\begin{eqnarray}
P_\mathrm{{pg}}(d) &=&  -k_\mathrm{B} T \partial_d \sideset{}{'}\sum_{l=0}^{\infty} {\rm Tr} \ln [ 1 - {\cal R}_\mathrm{{p}}(i \xi_l) 
\cdot  \chi_\mathrm{{pg}}(d,i \xi_l)  \nonumber \\
&&  \cdot 
{\cal R}_\mathrm{{g}}(i \xi_l) \cdot  \chi_\mathrm{{gp}}(d,i \xi_l) ]  .
\label{equation2}
\end{eqnarray}
Here $T$ is temperature,  ${\cal R}_\mathrm{{p(g)}}$ is the reflection operator of the plane (grating), and $\chi_{ij}$ are plane-wave translation operators between the two surfaces. The primed sum means that the $l=0$ term is counted with half weight. The trace operation sums over the two light polarizations, over different Brillouin zones of the periodic structure, and integrates over the parallel wavevectors $k_x$  (direction of grating modulation) and  $k_y$ (invariant direction for grating) from $-\pi/p$  to $\pi/p$   and  $-\infty$  to $\infty$, respectively. The broadband nature of the Casimir interaction is apparent in the above expression. The reflection operators are computed from the solution to Maxwell equations for the EM field conveniently decomposed in terms of the natural modes of the structures \cite{24}. In our calculation the temperature is set to $T = 300$ K and for simplicity we model the permittivity of gold using a Drude model  $\epsilon(\omega)=1-\Omega_\mathrm{p}^2/(\omega^2+i \Gamma \omega)$ with plasma frequency $\Omega_\mathrm{p}=8.39$  eV and dissipation rate $\Gamma=0.0434$ eV. Numerically, the errors in the computation of the Casimir pressure mainly arise from the truncation of the Matsubara sum and of the reflection and translation operators, represented as finite-size matrices. In our implementation, the total theoretical/numerical error is less than 2 \% over the entire pressure-displacement curve. The accuracy of our numerical results could nevertheless be affected by the model and optical parameters chosen to describe the actual permittivity of gold used in our samples. For example, a more accurate description of the optical properties for gold requires the introduction of the interband electronic transitions in the dielectric function. Generally, a good description of this effect is given by using the so-called 6-oscillator Drude-Lorentz model. We have checked that the inclusion of the interband contribution results in a plane-grating pressure a few percent stronger than the one given by the simple Drude model, and that at large distances its effect is negligible since the low frequency behavior is dominated by the Drude contribution. On the other hand, the use of a plasma model - $\Gamma=0$ - produces similar numerical results for the pressure in the whole experimental range.

The solid lines in Figure 2B and 2C are the result of the modal approach numerics for the equivalent plane-grating pressure $P_\mathrm{{pg}}$. At short distances this approach agrees with the data better than PFA, as was already observed in the case of silicon gratings \cite{30}. At large distances, however, we observe an even stronger disagreement with the experiment. By normalizing our numerical results obtain from Eq.(\ref{equation2}) by the values calculated using PFA given in Eq.(\ref{equation1}), the disagreement at large distance becomes even more evident (see Fig.3). In contradiction with the experiment, the two solid lines describing this ratio have values always larger than 1, attaining the maximum at distances of the order of the grating period. The ratio tends to 1 in two opposite limits: at short distance, substantially confirming the validity of the PFA (see Fig. 5 of the Supplementary Information) and at large distance, where both the equivalent plane-grating PFA pressure and the pressure calculated using the previous numerical approach tend to the same value, i.e. the Lifshitz plane-plane formula (see details in Methods). Several checks, including a comparison with the Casimir plane-grating pressure computed within the framework of an effective medium approach \cite{31}, have confirmed the validity of our calculation for the plane-grating geometry.

While disagreement with the usual PFA, Eq.(\ref{equation1}), was expected, the one with its modified version $2 \pi R P_\mathrm{{pg}}(d)$, using an exact theoretical description for the equivalent plane-grating interaction, is unanticipated. Two main differences distinguish our numerical calculation and the experimental setup: the PFA treatment of the sphere's curvature, i.e. the calculation of the equivalent plane-grating pressure, and the assumption of an infinitely periodic system, which is in contrast with the finiteness of all bodies used in the experiment. Both assumptions rely on common approximations that are known for providing a good theory-experiment agreement for planar unstructured samples, as long as $d$ is much smaller than $R$ and any lateral dimensions of the sample. As a check of the sensitivity of our experiment we have performed Casimir measurements with the sphere on top of the flat metallic pads, which have lateral dimensions similar to the grating (see Supplementary Fig. S4). 
We found good agreement between the data and the standard theoretical treatment that uses PFA, neglecting any finite-size effects of the pads, as well as an excellent agreement with the previous measurements of the Casimir force in the sphere-plane configuration. In this case deviations due to the curvature of our sphere tend to reduce the force with respect to its PFA value with a difference of less than 1 \% for distances shorter than one micron \cite{29}. 
The nanostructuring of the metallic surface, however, introduces into the problem additional length-scales and substantially modifies
the mode structure between the two plates.  In contrast to previous experiments \cite{20}, here we explored distances larger than the grating's period, where we observed the strong deviations from the theoretical values for the effective 
plane-grating pressure. Unfortunately, state-of-the-art numerical techniques cannot solve exactly the sphere-grating problem for our case given the disparate ranges of length-scales present in the experiment (100's nm-sized grating features, $>100 \; \mu$m sized sphere, and $< 1 \; \mu$m separation distances), preventing at the moment an  in-depth study of the validity of a PFA treatment of the sphere's curvature in our experimental sphere-grating geometry. 

In conclusion, we have shown that by nanostructuring the metal surface of interacting bodies at scales below the plasma wavelength, a new regime in the Casimir interaction can be achieved. 
This regime is significantly different from the well-known attraction between parallel plates and is characterized by a 
crossover from enhancement to strong reduction of the Casimir force. For large inter-surfaces separation, the Casimir interaction decreases faster than the usual $d^{-4}$ power law, reaching values more than $2$ times smaller than the one predicted by the proximity force approximation for planar-like geometries. 
We demonstrated that existing state-of-the-art theoretical modeling, based on the proximity force approximation for
treating the curvature of our large-radius sphere and an exact ab-initio scattering analysis of the resulting effective plane-grating geometry, does not reproduce the experimental findings. The development of a full numerical analysis of the sphere-grating
geometry, capable of dealing with the disparate length scales present in our experiment, remains an open problem.

\section*{Methods}

\noindent {\bf Sample fabrication.} The nanostructured gratings used in this work were fabricated using two different procedures as described below. \\

\renewcommand{\figurename}{{\bf Supplementary Figure S\hspace{-0.1cm}}}
\setcounter{figure}{0}

\begin{figure}[t]
\includegraphics[width=0.8\columnwidth,angle=0]{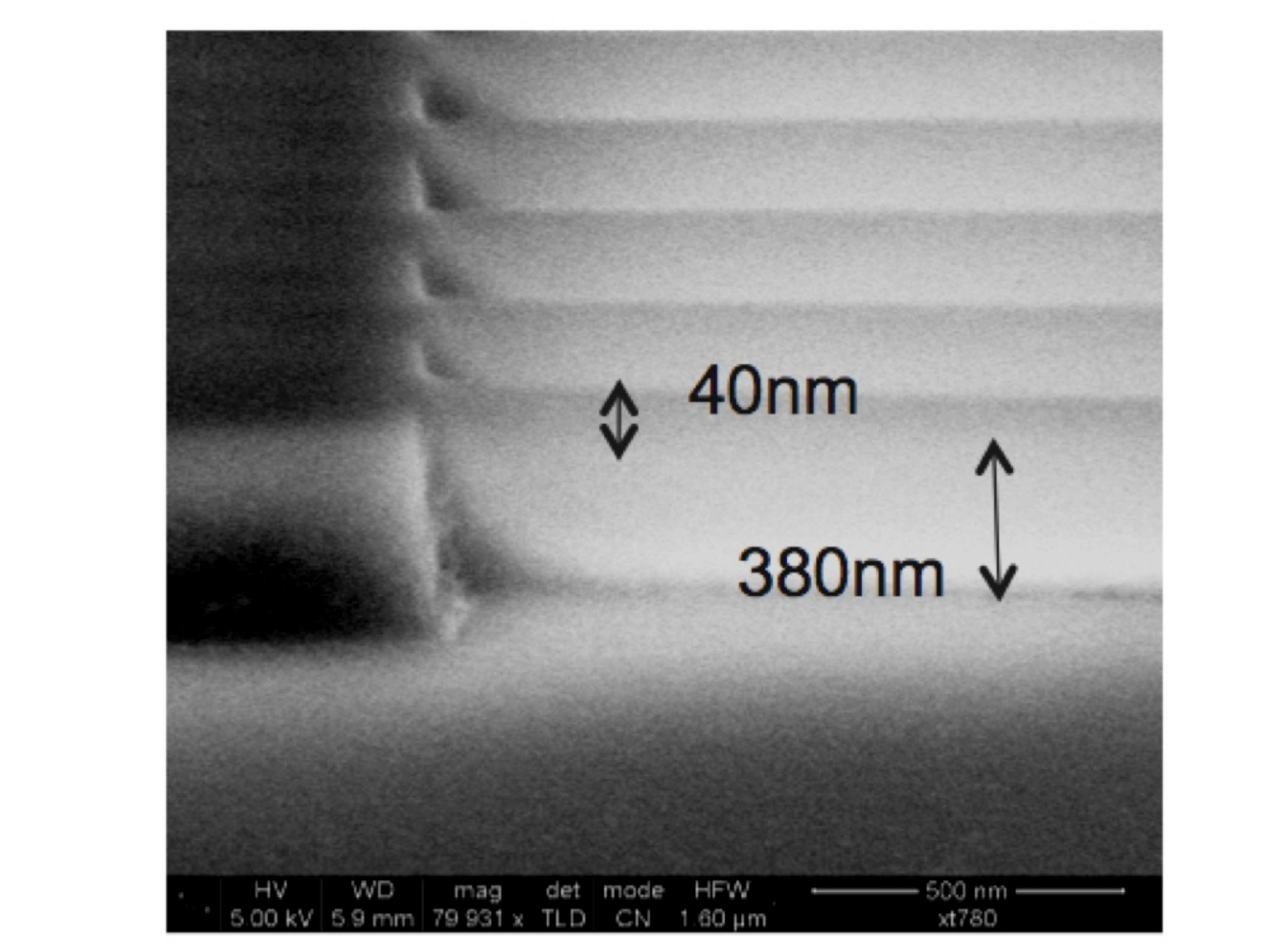}
\caption{ SEM image of a typical HSQ mold before deposition of Au.}
\label{SF1}
\end{figure}

\noindent {\bf Sputtering of gold onto hydrogen silsesquioxane (HSQ) structures.} 
HSQ is an inorganic negative tone e-beam resist which is basically a spin-on dielectric with silicon dioxide-like physical properties. HSQ was patterned with an e-beam lithography system, developed and cured.  In Supplementary Figure S1, we show a HSQ grating with lines having a height of 380 nm and width of 40 nm  ($\approx 10:1$ aspect ratio structure).
After patterning, these HSQ structures were coated with Au by conformal sputter deposition. The conformality of our deposition is around 0.25 and hence the widening of the lines is minimal but thick enough to completely cover the dielectric with Au. After deposition of 130 nm of metal, a 40 nm wide line becomes $a \approx 100$ nm wide line. This results in $4:1$ aspect ratio metallic nanostructures. This method does have its limitations in the smallest width achievable due to the minimal thickness required for Au in Casimir measurements. Since HSQ itself has very smooth surfaces after patterning it results in a smooth Au surface when coated. The Au surface quality from atomic force microscopy (AFM) measurements show that the deposited Au has a RMS surface roughness of $\approx 1.0$ nm and is comparable to the surface roughness of Au deposited on single crystal silicon. Varying structures with different widths (100 nm to 200 nm) and period (300 nm to 800 nm) were fabricated so that structures can be compared in measurements with different dimensions but similar filling factors. \\

\noindent{\bf Electroplating of high aspect ratio molds.}
In this method the gold gratings are fabricated by electron beam lithography and electroplating. Briefly, a very high-resolution positive e-beam resist (ZEP520) with thickness 500 nm is spun on Si chips coated with 5 nm Ti, 200 nm Au, and 5 nm Ti layers. The resist is exposed in an e-beam lithography system at 100 kV and developed in hexyl acetate to form a high-resolution, high aspect ratio mold. The top Ti layer is then etched off in reactive ion etching, so that the Au is exposed at the bottom of the mold (the top Ti layer is necessary due to the poor adhesion of the e-beam resist to the Au surface). Next, Au electroplating is performed in an AuCN bath at room temperature using a current density of 30 ${\rm A/m}^{2}$.  The deposition rate is measured experimentally, and the time is varied to attain the desired Au thickness. After electroplating, the chip is rinsed and the e-beam resist stripped in a solvent bath. The plated structures are characterized by optical profilometry, SEM, and AFM.  The measured RMS roughness of the lines is on the order of 1.5 nm, and the thickness variation across a structure is approximately 10 nm.  \\

\noindent{\bf Experimental setup and calibration.} 
We use a torsional oscillator (Fig. 1A) to measure the Casimir force between the gold sphere and the nanostructured gratings. The oscillator plate and a sapphire sphere are coated with a $\approx$ 1 nm layer of Cr followed by a $\approx$ 200 nm thick layer of Au. The oscillator is a 500 x 500 $\mu{\rm m}^{2}$, 3.5 $\mu{\rm m}$ thick heavily doped polysilicon plate suspended at two opposite points by serpentine springs. Serpentine springs were selected over conventional torsional rods because for equal sensitivity, they occupy a smaller region and reduce vertical sag of the torsional paddle. The springs are anchored to a silicon nitride covered Si platform. When no net torque is applied, the plate is separated from the platform by a $\approx$ 2 $\mu{\rm m}$ gap. Two independently contacted polysilicon electrodes located under the plate are used to measure the capacitance (Andeen-Hagerling AH2700A capacitance bridge) between the electrodes and the plate. The oscillator/sphere assembly is mounted on a 5-axis stepper motor driven positioner (Newport 561 series). The nanostructured surface is mounted on a $xyz$ piezo-driven, closed-loop, 70 $\mu{\rm m}$ range per axis (MadCity Labs Nanopositioning System). Both positioning systems, which are attached to a rigid, 5 kg stainless steel structure, allow for positioning and repeatability better than 0.2 nm. The whole assembly is contained in a vacuum chamber maintained at $P = 2.6 \times 10^{-5}$ Pa. There is passive magnetic damping between the assembly and the vacuum chamber. The vacuum chamber is mounted on an optical table with active vibration isolation control (TMC Precision Electronic Positioning System). As measured at the sample's position, the vibrational amplitude of motion is smaller than 10 pm in the 10 Hz to 1000 Hz range. The sphere used in the experiments has a radius of curvature $R = (151.7 \pm 0.2) \; \mu{\rm m}$.  The physical parameters for the sphere (radius and sphericity) were determined by means of SEM. Both were found to be within the specifications of the manufacturer. Deposition induced asymmetries were found to be smaller than 10 nm, the resolution of the SEM.  A single mode optical fiber (Corning SM-28) is rigidly attached to the nanostructured grating, and it is used to constantly monitor the absolute separation $D$ between the end of the fiber and the substrate below the torsional oscillator. The RMS error in the interferometric measurements is $\delta D$ = 0.25 nm, dominated by the overall stability of the closed-loop feedback system. Details on how the separation $d$ between the sphere and a uniform sample (i.e. the pad in Fig. 1B) is obtained can be found in previous work \cite{33}. 

The apparatus is calibrated using the electrostatic interaction between the sphere and the grating/pad plate. The torsional spring constant $\kappa =$ (8.85 $\pm$ 0.03) Nm is found in this way. Once the system is characterized a potential difference $V_{0}\not= 0$ between the sphere and the pad is applied to minimize, within the experimental error, the electrostatic force. We checked that $V_{0}$ is independent of position when the sphere is either on top of the pad or above the nanostructured surface.

\begin{figure}[t]
\centering
\includegraphics[width=0.9\columnwidth,angle=0]{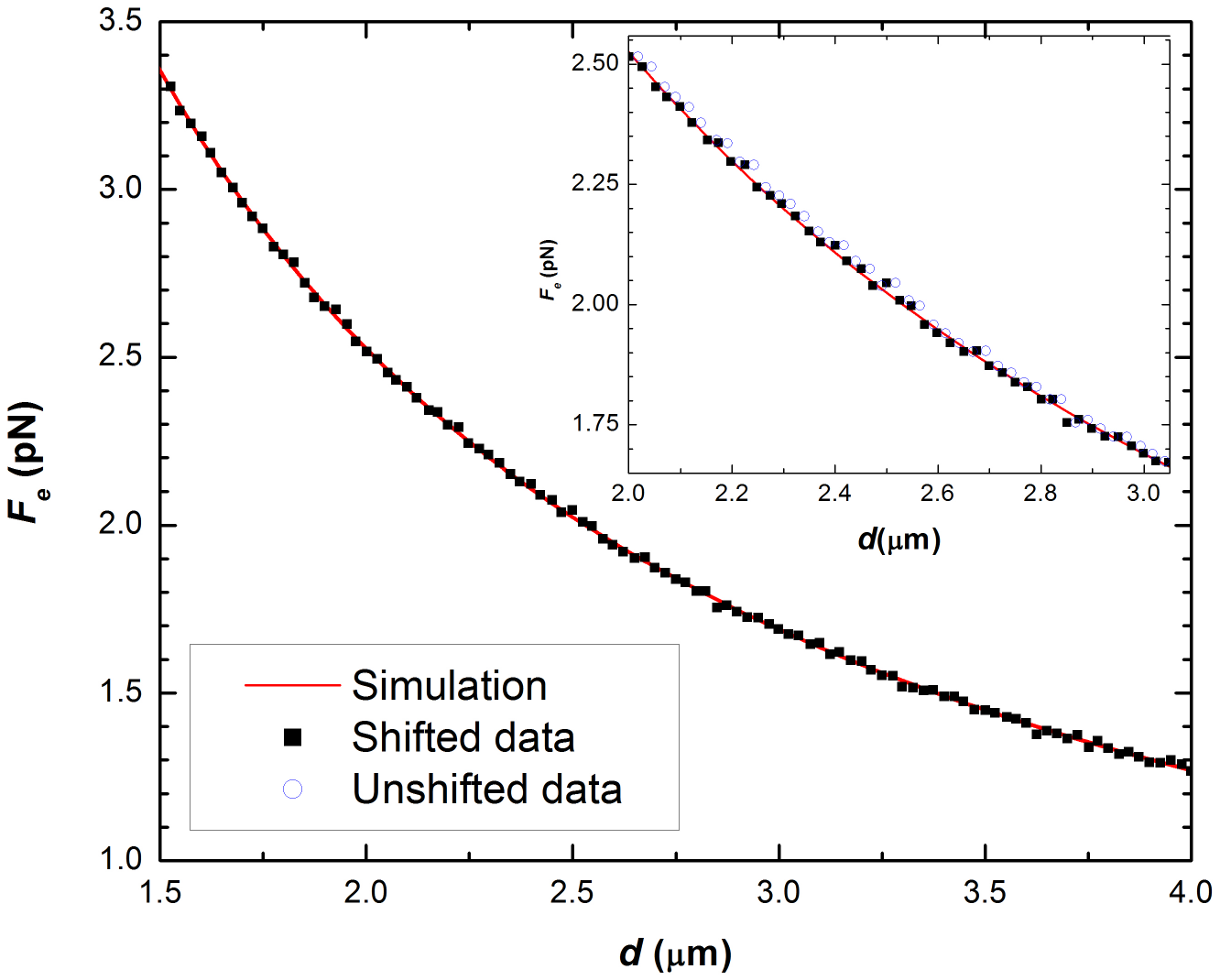}
\caption{Plane-grating electrostatic force obtained using a commercial finite element electrostatic solver (solid lines) and the shifted measured data (black squares). The inset shows, in addition to these two sets, the data before shifting (open circles).}
\label{SF2}
\end{figure}

Samples grown by metal sputtering have equal pad and grating heights, as determined by AFM measurements. Consequently the distance $D$ measured between the fiber and the pad was used for obtaining the distance $d$ between the apex of the sphere and the nanostructured surface. In contrast, in samples grown by electroplating  the pad is not as high as the grating, the height difference depending on the preparation conditions. In this case, two different approaches were used.

In the first approach, the capacitance and the electrostatic force between the sphere and the grating were measured as a function of separation $d$. These values were compared with calculations performed using a commercial finite element electrostatic solver.  To this aim we first calculated the capacitance per unit of area between a plane and a grating by numerically solving the electrostatics equations. The modeling was conducted using Comsol 3.51 software. Translational symmetry in the direction of the grating lines allows us to formulate the problem in 2D, while the periodicity in principle allows us to model only one period. In practice the model included two periods, because if was computationally affordable and provided for better visualization. Parametric study as a function of gap was performed by smoothly deforming the mesh without re-meshing for each gap to avoid numerical noise in the capacitance derivative. Two models were created: One with a bigger domain suitable  for modeling gaps above 500 nm, while the other with a smaller domain was used for the gaps in the 100 nm to 600 nm range.
For each model a mesh was created with higher density of elements near the ends of grating fingers. These meshes were further refined twice and the modeling repeated for each refinement. The refined meshes have 164460 (137768) degrees of freedom, 15093 (13371) mesh points and 29696 (26288) triangular elements for the larger (smaller) model domains. Typical relative numerical error in the capacitance values between the two models and the three different meshes within each of the models is below $2 \times 10^{-3}$ with better agreement between the finer mesh cases.  This imprecision is significantly below the other sources of uncertainly in the experiment. The metal surfaces were assumed perfect conductors at fixed electrical potentials. To calculate the capacitance as a function of separation distance, the mesh was smoothly deformed as the plane - grating separation was changed. Solutions at different mesh densities were compared to ensure numerical accuracy. Finally, PFA was used to obtain the gradient of the capacitance for the sphere/grating configuration, $\partial _d C$. Since the measured values of capacitance are inherently affected by parasitic capacitance, the comparison with the electrostatic force was deemed more reliable. In this scenario the gradient of the capacitance with respect to the separation was used to determine the calculated electrostatic force $F_{\rm e} =1/2 \partial _d C \Delta V^2$, where $\Delta V$ is the potential difference between the two plates of the capacitor. It was observed that the experimental curve, when the data was plotted as a function of $d$ (when the sphere is on top of the grating) has to be shifted by a sample-dependent amount $d_{0}^{i}$ ($\approx$ 15 nm) to make the calculated and measured values coincide. 

In the second approach, the height difference $d_{0}^{ii}$ between the pad and the ridges was measured using an AFM. This difference was taken into account when determining the separation between the sphere and the nanostructured surface.  Specifically, for the sample with $h$ = 400 nm, $w = 130$ nm, $p$ = 350 nm, $d_{0}^{i}$ = (17 $\pm$ 2) nm was obtained by a least square fitting of the electrostatic force with a single fitting parameter, as shown in Sup. Fig. 2. Using the AFM method, $d_{0}^{ii}$ = (17 $\pm$ 1) nm was found. In the latter case the error is the standard deviation of the values found when measuring the height difference at 25 different points. Since both methods yielded identical results within the experimental error, method (ii) was preferred due to its smaller intrinsic error.

\begin{figure}[t]
\centering
\includegraphics[width=0.9\columnwidth,angle=0]{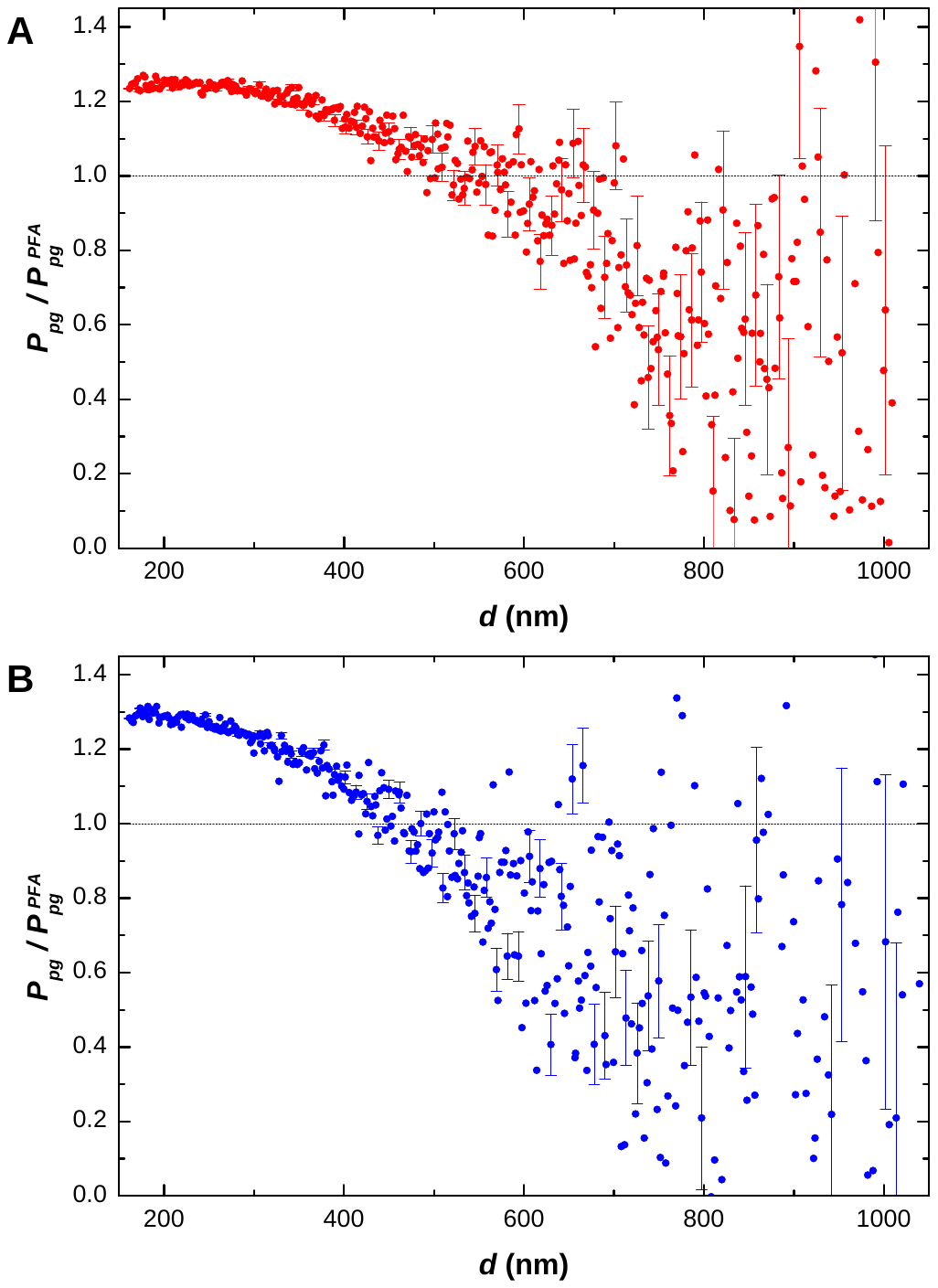}
\caption{Equivalent plane-grating Casimir pressure normalized by the PFA expression
shown in Eq. (1), as a function of separation between the sphere and the electroplated gratings. The upper panel shows the results obtained using sample 2 and the lower panel the ones from sample 1. Since the calculation of $P^{\rm PFA}$ is assumed to be exact, error bars are the variance of the mean measured pressure over the 45 repetitions of the experiment for each sample. They are plotted every fifth data point to
increase the clarity of the figure.}
\label{SF3}
\end{figure}

\begin{figure}[t]
\centering
\includegraphics[width=1.0\columnwidth,angle=0]{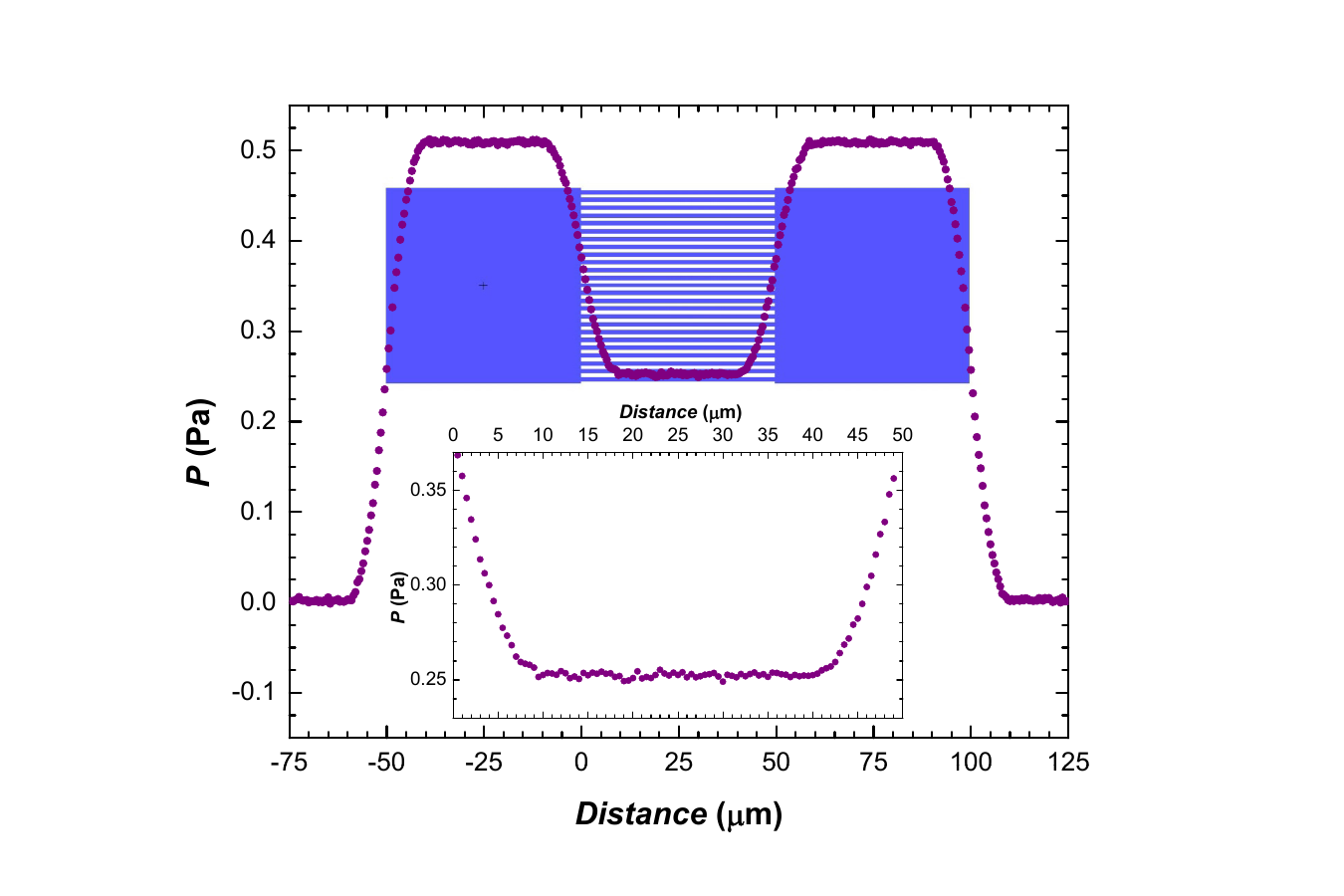}
\caption{Equivalent pressure as a function of position 
when the sphere is scanned on top of the pad-grating system. Data was 
acquired at a sphere-grating separation of 200 nm. The sample used was 
Sample 3 (300/130/480). For clarity, the schematic of the pad-grating assembly is shown. Inset: Zoom in on the grating region, showing the flatness of the data on the center region.}
\label{SF4}
\end{figure}

Data reported in Fig. 3  were obtained by performing a weighted rolling average over $n$ consecutive points. The value of the pressure at position $\overline{d}$ is
\begin{equation}
\overline{P}_\mathrm{{pg}}(\overline{d})=\frac{\sum_{i=0}^{n-1}P_\mathrm{{pg}}(d+i\Delta_\mathrm{{d}}d)\delta P^{-2}_\mathrm{{pg}}(d+i\Delta_\mathrm{{d}}d)}{\sum_{i=0}^{n-1}\delta P^{-2}_\mathrm{{pg}}(d+i\Delta_\mathrm{{d}}d)},
\end{equation}
where $\overline{d}=(1/n)\sum_{i=0}^{n}(d+i\Delta_\mathrm{{d}}d)$ , and $d+i\Delta_\mathrm{{d}}d$  represent the $n$ different separations considered. $\delta P_\mathrm{{pg}}$  is the random error in the determination of the pressure $P_\mathrm{{pg}}$  at distance $d$. The random error of the weighted rolling average is $\delta\overline{P}_\mathrm{{pg}}(\overline{d})=\left[\sum_{i=0}^{n-1}\delta P^{-2}_\mathrm{{pg}}  (d+i\Delta_\mathrm{d}d)\right]^{-1/2}$.  The number of data points $n$ used in the rolling average varies as a function of separation: $n = $10 for $d < 300$ nm and then it increases linearly with separation to reach a value of $n =$ 35 at $d =$ 1000 nm, the maximum separation between the sphere and the grating. The total error in the pressure is obtained as the addition of the systematic and random errors. The maximum contribution to the systematic error $\delta \overline{P}^{\rm syst}_\mathrm{{pg}}$  arises from the uncertainty in the measurement of the resonant frequency ($\delta f_{r}=$ 6  mHz) and from the error in the measurement of the sphere's radius $R$. The systematic error is smaller than the random error in the whole separation range. Between 300 nm $< d< $1000 nm,  $\delta \overline{P}^{\rm syst}_\mathrm{{pg}}\approx$ 0.2 mPa. In the binning process, the error in the separation is determined as the variance of the different separations used, which is dominant when compared to the error in the measurement of the separation  $\delta d\approx$ 2 nm. The experimental data before this smoothing procedure are shown in Supplementary Figure S3.

The measurements reported in Fig. 3 and in Supplementary Figure S3, were performed at the center of each grating. In order to find 
the center of the grating the sphere was scanned on top of the sample 
until a region where the measured signal did not depend upon position was 
found. One of such scans is shown in Supplementary Figure S4. This scan was 
performed in 0.5 $\mu$m intervals along the long axis of the pad-grating 
system. Outside the pads (left and right parts of the figure) the 
separation between the sphere and the sample is large and as a consequence, the signal 
is very small. For the measurements performed on top of the pads and on top of the grating, we observed 
that there is a finite region (about 30 $\mu$m across) where the 
signal is independent of the position.\\

\noindent{\bf Theoretical and numerical methods.}
Within the scattering approach to Casimir physics, the calculation of the plane-grating Casimir pressure is essentially reduced to the computation of the reflection operators  $\mathcal{R}$ of the plane and the grating. For the plane $\mathcal{R}_\mathrm{{p}}$ is given by the usual Fresnel coefficients. For the grating  $\mathcal{R}_\mathrm{{g}}$ is computed following the modal approach \cite{24}. We divide the grating geometry into three regions (see inset of Fig. 1A): (1) the vacuum, homogeneous region $z>h$  above the grating, (2) the grating region $0\le z\le h$, periodically modulated along the $x$-direction and invariant along the $y$-direction, and (3) the Au bulk, homogeneous region $z<0$  below the grating. Within each $i$-th region the EM field can be expressed as a series in terms of the eigenvectors which are solutions to Maxwell equations, namely
\begin{equation}
F^{(i)}(x,y,z,t)=\sum_{m}A^{(i)}_{m}Y^{(i)}_{m}(x)e^{i \lambda^{(i)}_{m}z}e^{i(k_{y}y-\omega t)}.
\end{equation}
Here $F$ denotes any component of the electric or magnetic field, and the sum is over a discrete set of complex eigenvalues  $\lambda^{(i)}_{m}$; the corresponding eigenvectors are denoted by $Y^{(i)}_{m}$. These quantities are computed using the quasi analytical approach discussed in the second reference in \cite{23}.  The complex coefficients $A^{(i)}_{m}$ are then determined by imposing boundary conditions on the vacuum-grating and grating-bulk interfaces, and finally the reflection operator $\mathcal{R}_{\rm g}$  is obtained and employed in the Matsubara series expression for the Casimir pressure.

Some analytical predictions can be made about the plane-grating pressure. First, at very large separations it is dominated by the low-frequency/low-momentum behavior of the reflection matrices: above $\approx$ 3 $\mu{\rm m}$ the zeroth Matsubara term is practically describing the whole interaction. For this term, using the Drude model for Au it is also possible to analytically solve for the eigenvalues and the eigenvectors of the EM field expansion in the grating \cite{34}. Only the transverse magnetic components matter, and in this limit the corresponding reflection matrices are equal to unity, for both the plane and the grating (the latter fact was numerically verified). Since 
the PFA expression for the plane-grating pressure, 
\begin{equation}
P_\mathrm{{pg}}^{\rm PFA}=f  P_\mathrm{{pp}}(d) + (1-f) P_\mathrm{{pp}}(d+h) ,
\end{equation}
shows exactly the same behavior in the same distance range, the ratio $P_\mathrm{{pg}}/P^{\rm PFA}_\mathrm{{pg}}$  must go to 1 at separations much larger than the ones accessed in the experiment (see Supplementary Figure S3).

\begin{figure}[t]
\centering
\includegraphics[width=0.9\columnwidth,angle=0]{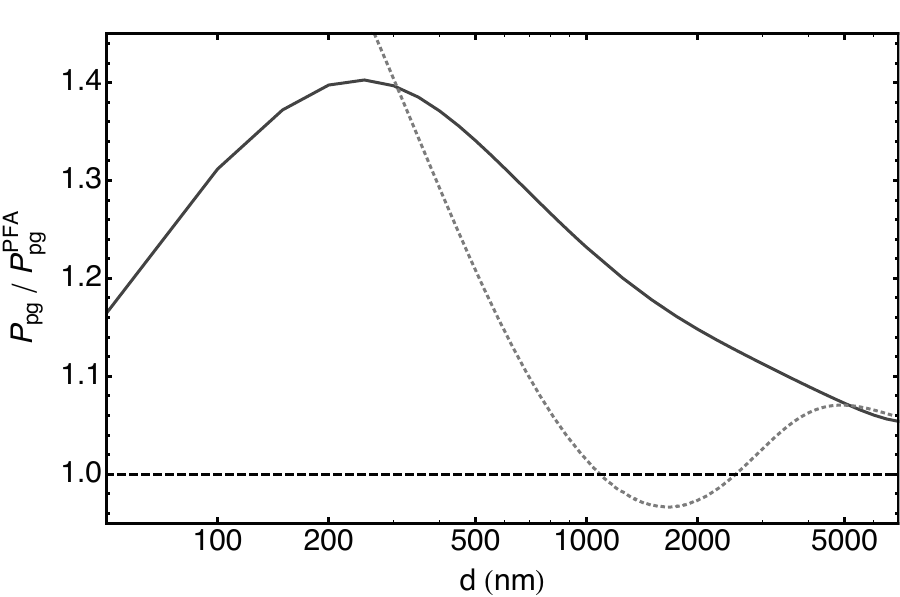}
\caption{ Plot of the equivalent plane-grating pressure for sample 1 normalized by the corresponding PFA prediction in Eq.(1). The solid line is the full numerical result for the modal approach as in Eq. (2), and the dotted line represents the result of the calculation performed within an effective medium approach (see text). These results show that effective medium cannot be trusted below 5 $\mu$m.}
\label{SF5}
\end{figure}

Second, in order to gain further insights into the large-separation behavior, we calculated the plane-grating Casimir pressure using an effective medium approximation  (EMA) for the nanostructure \cite{35}. This approximation consists in replacing the spatial-dependent electric permittivity $\epsilon(\omega,\mathbf{r})$ describing the geometrical and optical properties of the nanostructure by an effective homogeneous (not necessarily isotropic) permittivity, $\overleftrightarrow{\epsilon}_{\rm EMA}$. The EMA is expected to be valid for separations much larger than the geometrical features of the nanostructure (above $\approx$ 5 $\mu$m). The EMA permittivity tensor is modeled as that for a uni-axial anisotropic medium \cite{28},  $\overleftrightarrow{\epsilon}_{\rm EMA}=\mathrm{diag}(\epsilon_{xx},\epsilon_{yy},\epsilon_{zz})$ with $\epsilon_{yy}=\epsilon_{zz}=\epsilon_{D}f+(1-f)$ and $\epsilon_{xx}=\epsilon_{D}\left[f+\epsilon_{D}(1-f)\right]^{-1}$, and the resulting EMA plane-grating pressure is calculated following the technique of previous work \cite{31}. As follows from Supplementary Figure S5, the EMA fails to reproduce the exact results for distances below 5 $\mu$m. In the expected range of validity of EMA, the ratio  $P^{\rm EMA}_\mathrm{{pg}}/P^{\rm PFA}_\mathrm{{pg}}$  is always close to 1.

Third, at short ($d\le$ 400 nm)  and intermediate (400 nm $\le d\le$ 1000 nm)  separations, the respective enhancement and reduction of $P_\mathrm{{pg}}/P^{\rm PFA}_\mathrm{{pg}}$  are stronger for the grating with the shorter period (see Fig. 3). It should be noted that the two fabricated samples have slightly dissimilar filling factors. However, it can be numerically shown that for gratings with \emph{identical} filling factors, an analogous behavior occurs. This feature can be understood with the help of the following scaling argument. In connection with the scale invariance of Maxwell equations, the plane-grating Casimir pressure satisfies the scaling property  $P_\mathrm{{pg}}(d,p,f,l)=p^{-4}P_\mathrm{{pg}}(d/p,1,f,l/p)$, where $l$ denotes all other characteristic lengths in the problem (height, Au plasma wavelength, thermal wavelength, etc.). Let us suppose that in the distance regimes considered,   $P_{\rm pg}$ does not appreciably depend on $l$. Then, for two gratings with identical filling factors ($f=f_{1}=f_{2}$)  but different periods ($p_{1}<p_{2}$), the respective pressures are given by $P_{1}\approx p_{1}^{-4}P(d_{1}/p_{1},1,f)$ and $P_{2}\approx p_{2}^{-4}P(d_{2}/p_{2},1,f)$. One expression can be obtained from the other by using the set of linear transformations: $P_{2}=(p_{1}/p_{2})^{4}P_{1}$ and $d_{2}=(p_{1}/p_{2})d_{1}$ . As a direct consequence, if the pressure can be approximately described by a power-law $P\propto d^{-n}$ in a certain region of distances, then $P_{1}\propto d^{-n}$  implies  $P_{2}\propto (p_{1}/p_{2})^{4-n}d^{-n}$. Therefore, at the same plane-grating distance,  $P_{2}<P_{1}$  for  $n<4$ (this is the scenario in Fig. 3 at short separations) and  $P_{2}>P_{1}$  for  $n>4$ (this is the scenario in Fig. 3 at large separations). Small deviations from the above scaling argument are due to the role played by $l$, and also to the slightly different filling factors of the two fabricated samples. \\

\noindent{\bf Acknowledgements:} We are grateful to R. Behunin, H.B. Chan, J-J. Greffet, R. Gu\'erout, S. Johnson, S. de Man, P. Milonni, J. Pendry, F da Rosa, and T. Kenny for discussions. 
The full description of the procedures used in this paper requires the identification
of certain commercial products and their suppliers. The inclusion of such information should in
no way be construed as indicating that such products or suppliers are endorsed by NIST or are
recommended by NIST or that they are necessarily the best materials, instruments, software
or suppliers for the purposes described. This work was partially supported by the DARPA/MTO Casimir Effect Enhancement program under DOE/NNSA Contract No. DE-AC52-06NA25396 and DOE-DARPA MIPR 09-Y557. RSD acknowledges support from the IUPUI Nanoscale Imaging Center, Integrated Nanosystems Development Institute, Indiana University Collaborative Research Grants and the Indiana University Center for Space Symmetries.  This work was performed, in part, at the Center for Nanoscale Materials, a U.S. Department of Energy, Office of Science, Office of Basic Energy Sciences User Facility under Contract No. DE-AC02-06CH11357. \\

\noindent{\bf Additional information:} The authors declare no competing financial interests. Supplementary information accompanies this paper.\\

\noindent{\bf Author Contributions:} The theoretical work was carried out by F.I., P.D., V.A.A. and D.A.R.D. The experimental work was carried out by S.K., I.W.J., A.A.T., R.S.D., V.A.A. and D.L.\\

\noindent{\bf Author Information:} Correspondence and requests for materials should be addressed to D.L. (dlopez@anl.gov).


 \end{document}